\documentclass[fleqn,12pt,twoside]{article}
\usepackage{espcrc1}
\usepackage{epsf}

\def\LC{LC}

\def\be{\begin{equation}}
\def\ee{\end{equation}}
\def\bea{\begin{eqnarray}}
\def\eea{\end{eqnarray}}
\newcommand{\beq}{\begin{equation}}
\newcommand{\eq}{\end{equation}}
\newcommand{\rb}{\underline{r}}
\newcommand{\kb}{\underline{k}}

\newcommand{\ab}{\underline{a}}
\newcommand{\bb}{\underline{b}}
\newcommand{\pb}{\underline{p}}

\newcommand{\zb}{\bar{z}}

\newcommand{\intfeyn}{\int\limits_0^1}

\newcommand{\AmS}{{\protect\the\textfont2
  A\kern-.1667em\lower.5ex\hbox{M}\kern-.125emS}}

\title{$\gamma ^* \gamma ^* \to \rho \rho $ at very high energy.}

\author{B.~Pire\address{CPhT, \'Ecole Polytechnique, 
        91128 Palaiseau, France, UMR 7644 du CNRS},         
        L.~Szymanowski\address{Soltan Institute for Nuclear Studies, Ho\.{z}a 69, 00-681
Warsaw, Poland and
Universit\'e de Li\`ege, 
B4000 Li\`ege, Belgium},
   S.~Wallon\address{ LPT, Universit\'e d'Orsay,
   F 91405-Orsay, France, UMR 8627 du CNRS}     }
       
\begin{document}

\maketitle

\begin{abstract}
The next generation of $e^+e^--$colliders will offer a possibility of
clean testing of QCD dynamics in the Regge limit. Recent progress in the theoretical
description of exclusive processes permits
for many of them a consistent use of the perturbative QCD methods.
 We advocate that  the exclusive
diffractive production of two $\rho$ mesons
from virtual photons at very high energies should be measurable at the future
linear collider (LC).
\end{abstract}

\section{Introduction}

The high energy limit of strong interaction has a very long story, which 
started much before the development of QCD \cite{revue}. The Regge {\it limit} corresponds to 
the kinematical regime of
large scattering energy square $s$ and small momentum transfer square $t$,
$s \gg -t$.
Soon after QCD was proposed as a theory for strong interactions,
its Regge  limit was studied by Balitsky, Fadin, Kuraev and 
Lipatov \cite{bfkl}.
The evaluation of the elastic scattering amplitude of two 
infrared safe objects was performed, as an infinite series in $\alpha_s
\, ln s\,.$ This so-called Leading Log Approximation (LLA), where small values
 of perturbative $\alpha_s$  are compensated by large values of $\ln s,$
is expressed as an effective ladder with two reggeized gluons in
$t$-channel (gluons dressed by interaction, 
resulting in appearence of Regge trajectories) interacting with
$s-$channel
gluonic rungs,
through the effective Lipatov vertex which generalizes the usual
triple Yang-Mills vertex. The net result for this {\it hard} Pomeron intercept
is $\alpha_P(0)= 1+ c \, \alpha_s,$ where $c$ is a stricly positive constant,
which thus leads to a violation of the Froissart bound at perturbative
level.

In order to test the hard Pomeron, it is not enough to study large $s$
experiments. It is also compulsory to select processes where a hard
scale enables one to use perturbative QCD. 
In DIS, the virtuality of the photon naturaly provides a hard scale.
At the level of both total and diffractive cross-sections,
 it was possible to describe HERA data using models based on BFKL
type of evolution, although the distinction with  standard DGLAP
evolution \cite{dglap} is not conclusive \cite{F2}.
Exclusive vector meson production was also proposed in order
to see BFKL effects, selecting events with a large gap in rapidity
between the vector meson and the outgoing proton (or its remnants).
These approaches needed however some ansatz for the non-perturbative
proton-Pomeron coupling.

\section{$\gamma^*\gamma^*$ processes: the gold plated experiment}

From the theoretical point of view, 
the best way for studying typical Regge behaviour in perturbative
QCD is provided by the scattering of small transverse size objects. 
Such a reaction is naturally provided by photons of high
virtuality as produced in $e^+e^-$ tagged collisions.
This was investigated at the level of total  $\gamma^*
\gamma^*$
cross section  
by various groups \cite{bfklinc}. Typical Pomeron
enhancement can hardly be seen at LEP, but should be definitely measurable at
\LC.
One of the key point in order to reveal this effect is that the detectors
should be able to tag the outgoing particle with minimal tagging angle down to
20 mrad.

Another possibility is to select specific $heavy$ bounds states
($J/\Psi,\Upsilon,...$) in the final state. 
This has been studied in the case of double diffractive
photo production of $J/\Psi$ \cite{jpsi}.
Several tens of thousand events are expected at \LC,
with an enhancement factor of the order of 50 with respect to the Born 
estimate.

\begin{figure}[htp]
\epsfxsize=4.4cm{\centerline{\epsfbox{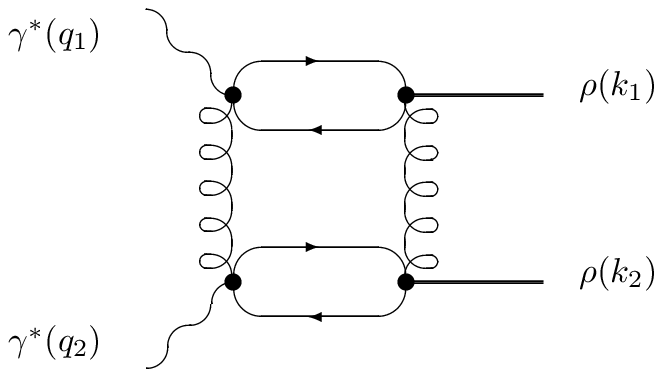}}}
\caption{Amplitude for the process $\gamma^*\gamma^* \to \rho\rho$
at Born order. The dots denote the effective coupling of t-channel gluons to
the impact factors. Virtualities are defined by $Q_{1(2)}^2=-q_{1(2)}^2.$}
\label{dessinprocess}
\end{figure}

We study the process of exclusive electroproduction
of two  $\rho-$mesons in $\gamma^* \gamma^*$
collisions. The measurable cross section in $e^+ e^-$ collisions is related to the amplitude of 
this process through the usual photon flux factors :
 \beq
\label{eesigma}
\frac{Q_1^2Q_{2}^2 d\sigma(e^+ e^- \to e^+ e^- \rho \rho)}{dy_{1}dy_{2}dQ_{1}^2dQ_{2}^2}
= \frac{\alpha}{2\pi} P_{\gamma/e}(y_{1}¥) P_{\gamma/e}(y_{2}¥) 
\sigma(\gamma^*\gamma^*\to \rho \rho)\,,
\ee
where $y_{i}$ are the longitudinal momentum fractions of the Brehmstrahlung photons with 
respect to the respective leptons and with $P_{\gamma/e}(y) = 2(1-y)/y$ for longitudinally 
polarized photons.
The virtualities  $Q_1^2$ and $Q_2^2$ of the scattered photons play
 the role of the hard scales. 
 This allows one to  scan $Q_1^2,$ $Q_2^2,$
as well as  $t$ to test the structure of the hard Pomeron. It is also possible to study various polarizations of both photons and
mesons.
As a first step in this direction we have calculated the Born order contribution to this
process with  longitudinally polarized photons and 
$\rho-$mesons, as illustrated in Fig.\ref{dessinprocess}.
 The choice of longitudinal polarizations of both the
scattered photons and produced vector mesons is dictated by the
fact that this configuration of the lowest twist-2 gives the dominant
contribution in the powers of the 
hard scales $Q_{1,2}^2$. $Q_1^2$ and $Q_2^2$ should be taken of the same order 
 so as to reduce phase space for conventional
parton evolution.

We use the usual impact representation, where the meson vertex is treated in the collinear 
approximation which neglects in the hard part of the amplitude the relative transverse
momentum of the quarks. This results in appearence of the
Distribution Amplitude (DA): the
meson wave function integrated over the relative momentum of quarks.

The amplitude for the process reads, defining $\zb =1-z,$
\beq
\label{MCalgeneral}
{\cal M} = -i\,  s \, 2 \pi \frac{N_c^2-1}{N_c^2} \, \alpha_s^2 \, \alpha_{em}
 f_\rho^2 \, Q_1 \, Q_2  \intfeyn d z_1 \, d z_2 \, z_1
\, \zb_1 \, \phi(z_1)\, z_2
\, \zb_2 \, \phi(z_2) {\rm M}(z_1,\, z_2)\,.
\eq
Here  $f_\rho$  is the $\rho-$meson coupling constant $\phi(z)=6 z \zb$ is the asymptotic DA of the $\rho$ meson, $z \,(\zb)$
being the light-cone quark (antiquark)  $\rho$ momentum fraction, and
\bea
\label{defM}
{\rm M}(z_1,\, z_2)=\int \frac{d^2 \kb}{\kb^2 (\rb-\kb)^2} &&\! \! \! \left[ \frac{1}{z_1^2\rb^2 + \mu_1^2} +
\frac{1}{\zb_1^2\rb^2 + \mu_1^2} 
 - \frac{1}{(z_1\rb -\kb)^2 + \mu_1^2} - \frac{1}{(\zb_1\rb
-\kb)^2
+ \mu_1^2} \right] \nonumber \\
&&\hspace{-4cm} \times \left[ \frac{1}{z_2^2\rb^2 + \mu_2^2} +
\frac{1}{\zb_2^2\rb^2 + \mu_2^2}
 - \frac{1}{(z_2\rb -\kb)^2 + \mu_2^2} - \frac{1}{(\zb_2\rb
-\kb)^2
+ \mu_2^2} \right] \! \! \! \! 
\eea
 is the transverse momentum convolution of the impact factors with 2 $t-$channel gluon propagators.
It can be expressed  in terms of three kind of integrals, namely

\beq
\label{I2}
I_2=\int \frac{d^d \kb}{\kb^2 (\kb - \pb)^2}\,, ~~~~~~~~~~~~~~~
I_{3m}=\int \frac{d^d \kb}{\kb^2 (\kb - \pb)^2((\kb-\ab)^2 + m^2)}\,,
\nonumber
\eq
\beq
\label{I4mm}
I_{4mm}=\int \frac{d^d \kb}{\kb^2 (\kb - \pb)^2((\kb-\ab)^2 + m_a^2)((\kb-\bb)^2 + m_b^2)}\,,
\nonumber
\eq
where we use the dimensional regularization $d = 2 +2 \epsilon$. 
 These  integrals were computed exactly using a generalized version of a technique used in 
 coordinate space when evaluating 
diagrams of massless two dimensional conformal field theories. 

\begin{figure}
\epsfxsize=6.4cm{\centerline{\epsfbox{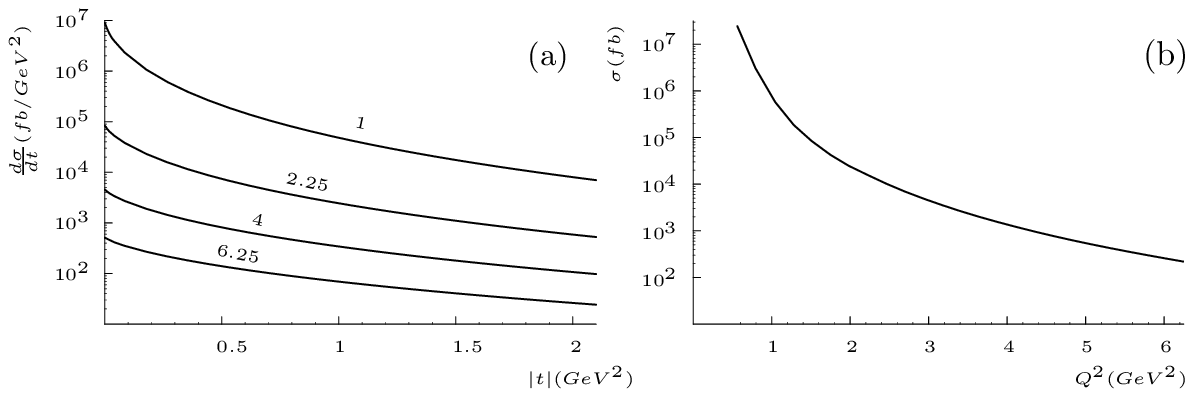}}}
\caption{Born order result for (a)\,${d \sigma^{\gamma^*\gamma^*\to \rho
  \rho}}/{dt}$ as a function of  $|t|\,(GeV^2)$ for various
  values of $Q^2\,(GeV^2)$, (b)\, $\sigma^{\gamma^*\gamma^*\to \rho \rho}$ as a function of $Q^2\,(GeV^2)$. }
\label{resultat}
\end{figure}

The final result for $M(z_1,z_2)$ is too lenghty to be given here.
It is regular in $z_1$ and $z_2.$ After numerical integration over $z_1$ and $z_2$ and squaring, one obtains the 
differential cross-section 
${d \sigma^{\gamma^*\gamma^*\to \rho \rho}}/{dt},$ shown 
in Fig.\ref{resultat}a,
 for various values of $Q_1^2=Q_2^2=Q^2.$
It is rapidly decreasing in $t$, and flat in $s.$ Any BFKL type of resummation
would give a rising shape in $s.$ 
Integrating over $t,$ one gets the $\sigma^{\gamma^*\gamma^*\to \rho \rho}$
cross-section, as shown in Fig.\ref{resultat}b.

The expected number of events at \LC, for a nominal luminosity of $100 fb^{-1},$ is of the order of 1000 events per year.
This is only a lower bound since the contribution of the transverse
photon case has to be added.
Morover, we expect  a net and visible enhancement of this cross section,
because of resummation effects \`a la BFKL. 

\section{Conclusions}

Double diffractive $\rho$ production in $e^+e^-$ collisions is 
a crucial test for QCD in Regge limit.
The Born contribution for longitudinally polarized photon and meson
 gives a measurable cross-section.
BFKL enhancement remains to be evaluated.

$e^+e^-$ collisions would be also a very good place to observe and test the Odderon. Such an object is the partner of the Pomeron, 
with opposite charge conjugation. We propose to study double diffractive $\pi^0$ production from two highly virtual photons,
which should be dominated by the $t-$channel exchange of an Odderon. In
QCD, such a state is constructed from at least 3 gluons, and resummation
effects are expected in the Regge limit \cite{ewerz}.
To test the existence of Odderon at the amplitude level,
one may study interference effects between Odderon and Pomeron exchange
in  $\gamma^*\gamma^*\to (\pi^+ \pi^-) (\pi^+ \pi^-)$ processes, using the
fact that the C-parity is not fixed for such final 
states \cite{pire}. 

\vspace*{.1cm}
\noindent {\bf Acknowledgments.}

\vspace*{.1cm}
\noindent    
This work is supported by the Polish Grant 1 P03B 028 28 and  by  the
French-Polish scientific agreement Polonium. L.Sz. is a Visiting Fellow
of the Fonds National pour la Recherche Scientifique (Belgium).

\end{document}